# Testing logic cores using a BIST P1500 compliant approach: a case of study

P. Bernardi[2], G. Masera[1], F. Quaglio[1], M. Sonza Reorda[2]

[1]Politecnico di Torino
Dipartimento di Elettronica
Torino, Italy

[2]Politecnico di Torino
Dipartimento di Automatica e Informatica
Torino, Italy

*Abstract*

*In this paper we describe how we applied a BIST-based approach to the test of a logic core to be included in System-on-a-chip (SoC) environments. The approach advantages are the ability to protect the core IP, the simple test interface (thanks also to the adoption of the P1500 standard), the possibility to run the test at-speed, the reduced test time, and the good diagnostic capabilities. The paper reports figures about the achieved fault coverage, the required area overhead, and the performance slow-down, and compares the figures with those for alternative approaches, such as those based on full scan and sequential ATPG.*

## 1. Introduction

Continuous technological improvements allow designing a complex system into a single chip (System-On-Chip or SoC). A SoC is composed of different reusable functional blocks, called embedded cores. Typical embedded cores include processors (such as CPUs, DSPs, and microcontrollers), memories (such as ROMs, DRAMs, SRAMs, and flash), I/O devices, etc. System designers can purchase cores from core vendors and integrate them with their own User-Defined Logic (UDL) to implements SoCs. Core-based SoCs present important advantages: the size and the cost of the end-product are decreased, and thanks to the design re-use, the time-to-market is greatly reduced.

Conversely, testing a core-based SoC is a major challenge [1]. The main problem is that accessibility to cores and UDL is greatly reduced. Traditional approaches [1-2] for testing core-based SoCs completely rely on additional Design for Testability (DfT) structures such as test busses for test transfers from/to the core under test. The access mechanism requires additional logic (such as a wrapper around the core) and wiring (such as a test access mechanism or TAM) to connect cores to the test source and sink. A critical point to be solved in SoC testing is the extra cost introduced by the DfT logic, i.e., the area, delay and test application time overheads. Some approaches have been proposed to solve this problem. A class of test approaches [3-4] adopts the reuse of existing functionalities for test access. These methods assume that every core has a transparent mode in which data can be propagated. However, these methods are not general enough to handle all the possible kinds of cores and all the possible test schemes, such as scan or Built-In Self-Test (BIST). Some researchers [5-7] proposed to exploit an embedded processor to test the other components of the SoC: first the processor core is tested (e.g., by means of functional patterns or full-scan test), and then a test program, executed by the embedded processor, is used to test the on-chip memories and other cores. The use of embedded processors to test cores presents many advantages: the size of the test controller is normally negligible, the test program (being in software) guarantees a high flexibility, and the testing process can often be done at-speed. Moreover, the test process is done inside the chip and the tester can work at a lower speed, thus reducing the costs for the test equipment. The main disadvantage is related to the need for an on-chip processor and to the dependence on the one possibly present in the SoC. In case of different processors, the test program has to be adapted to each of them causing an increased cost in the test development. Moreover, this approach cannot be applied to embedded cores not suitably connected to the processor.

When considering logic cores, the adoption of deterministic BIST is becoming increasingly popular, mainly due to the availability of efficient commercial tools supporting this technique. However, this approach has serious limitations when delay faults are of interest, since it relies on scan chains to access the internal points of the circuit.

In this paper we report the experience gathered by adopting a test strategy for logic cores based on a custom BIST engine wrapped to the embedded device under test. From the test point of view, the core complies with the P1500 standard, thus easing the connection of the core to other test resources on the chip (e.g., a 1149.1 TAP controller for accessing the SoC from outside). The advantages of this solution are first the high re-usability of the IP core, even in terms of testing features: only the test protocol must be delivered to the user (no vectors). As a second advantage there is the high fault achieved by the BIST engine (especially in terms of delay faults, as it performs the test at-speed). Finally, the approach provides the test engineer with some diagnostic feedback about the executed test; low area overhead and negligible costs in terms of wiring and performance reduction are the main drawbacks. The P1500 standard structure integrated with the BIST finally provides plug and play characteristics to the core to be easily inserted into powerful SoC level test structures. A case study will be reported allowing to better evaluating the advantages and disadvantages of the adopted test strategy.

Section 2 summarizes the possible architectural solutions for core-based SoC testing and introduces the one adopted in our work. Section 3 describes the proposed test architecture in terms of internal organization, detection and diagnosis properties evaluation, and external layers. Section 4 describes the considered case study and, finally, section 6 draws some conclusions.

## 2. Core test architectures

Testing core-based SoCs is a complex problem that can be divided in two parts: core-level and chip-level testing. *Core-level testing* involves making each core testable, i.e., inserting the necessary test structures and generating test sequences. *Chip-level testing* involves defining a test access mechanism architec-ture for applying the test sequences to the input peripheries of the cores, and for propagating the test responses from the core outputs to the chip outputs.

In this paper we mainly focus on the core-level testing problem, resorting to a custom TAM for connecting the core to a standard 1149.1 TAP controller, thus providing an easy interface between the SoC and an external ATE.

Different approaches can be adopted for testing logic cores; they are normally grouped in the following classes:
- scan based
- pseudo-random based [8-9]
- logic BIST.

In scan based and logic BIST approaches, a set of patterns are generated using automatic tools (ATPGs) and applied to the circuit. In the sequential approach, the calculated patterns are sequentially sent to the circuit and responses read after each application and any additional internal structure is added in order to improve the effectiveness of the patterns. On the contrary, in the scan approach, the controllability and the observability of the circuit are improved by modifying the common flip flop: the so-called scan cells allow writing and reading the content of the memory element during the test apply, and are connected to compose a scan chain. However, as a serial process is required to load and upload the scan chain, that approach requires onerous application time and heavy ATE requirements in terms of storage needed for test data and test application program.

An alternative technique is the pseudo-random pattern generation. Such approach is based on the Galois theories for the generation of pseudo-random number sequences starting from the definition of a characteristic polonium. Particular structures, called Autonomous Linear Feedback Shift Registers (ALFSR) provides a perfect hardware implementation for such kind of pattern generation strategy.

In this paper, a BIST implementation based on pseudo-random pattern generation is proposed and its effectiveness in coping with cores composed of many functional modules is underlined. In particular we focussed on the definition of a low-intrusive techniques guaranteeing high performance in term of fault coverage and high reusability in SoC structured. The proposed architecture is accessible through the test standard interfaces (IEEE 1149.1 and P1500) in order to make the test completely accessible from the outside: the test of the core can thus be executed by means of 1149.1 instructions and the core test details are completely transparent to core designers, maintaining the core Intellectual Properties.

## 3. The adopted approach

The approach adopted in this paper is based on the insertion of a BIST engine in charge of generate and apply patterns to the device under test. The adopted BIST circuitry has been designed to achieve two goals: on one hand we want to simplify the introduction of any modification in the pattern generation algorithm; on the other hand we aim at adopting the same architecture to a large set of logic cores. The adopted architecture exploits the current test standard interface in order to simplify the designer effort. The BIST engine is not directly visible from the outside, since the core is P1500 compliant, and includes a P1500 wrapper.

### 3.1. The BIST engine

The BIST engine internal architecture is divided into the following functional blocks:
- a **Control Unit** to manage the test execution;
- a **Pattern Generator** to produce and apply the test patterns;
- a **Result Collector** to manage the memory access timing.

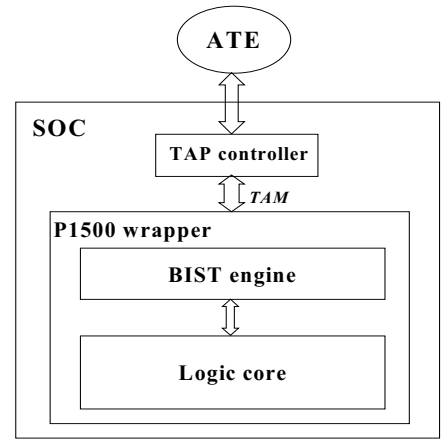

Fig. 1: The general architecture of the adopted approach.

In figure 2 a generic environment for the considered approach is presented: the A, B and C modules are parts of the same logic core and communicate among them in order to process inputs and generate outputs. In the picture, inputs and outputs of each module are considered separately to ease the test approach comprehension.

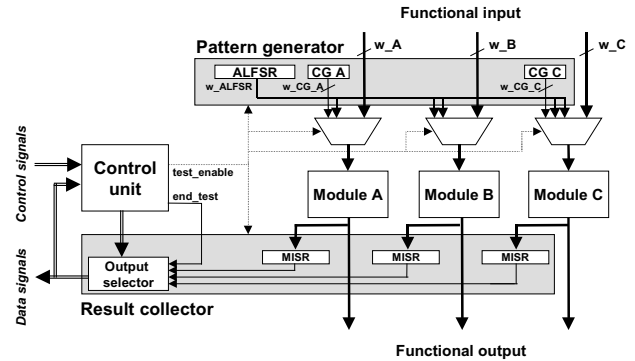

Fig. 2: BIST engine overall structure.

The **Control Unit** manages the test execution; by receiving and decoding commands from the control signals, this module is able to manage the test execution and the upload of the results. In particular, it covers three tasks:
- it receives from the data signals the number of patterns to be applied
- it drives the `test_enable` signal that starts and stops the test execution and provides the information about the end of the test
- it selects the result to be uploaded.

This choice allows easy reuse in different applications, like the application of test vectors generated according to different algorithms, or the test of logic cores with different characteristics.

The **Pattern Generator** is in charge of the application of the patterns to the DUT and is composed of:
- an ALFSR module[8-9]
- a set of Constraints Generators (CG).

The ALFSR module generates pseudo-random patterns according to the chosen polynomial characteristics; for cores composed of many functional blocks, only one ALFRS circuitry can be employed. On the contrary, a Constraint Generator is a custom circuitry able to drive constrained inputs. The adoption of such blocks provides great improvements in terms of effectiveness of the applied test, where a particular state machine controls the behavior of the circuit.

In the design of the pattern generation circuitry, we identified four architectural situations:
a. the block under test does not have constrained inputs and the ALFRS size fits the input port width
b. the block under test does not have constrained inputs and the input port width is larger than the ALFRS dimension
c. the block under test does have constrained inputs and the ALFRS size fits the input port width
d. the block under test have constrained inputs and the input port width is bigger than the ALFRS dimension.

While in a) the designer task consists just in connecting the ALFRS output with DUT input port, in b), c) and d) scenarios more care in choosing connections is required. Respectively, designers have to
- replicate the ALFRS outputs to reach the input port width
- identify the constrained inputs to build the CG and connect the ALFRS output to the remaining inputs.
- identify the constrained inputs to build the CG and replicate the ALFRS outputs to drive all the remaining inputs.

These three situations are shown in figure 2 where respectively $w\_B > w\_ALFSR$, $w\_A < w\_ALFSR + w\_CG\_A$ and $w\_C > w\_ALFSR + w\_CG\_C$.

The **Result Collector** is in charge of storing and making the results reachable from the outside. It is composed of
- a set of MISR modules[8-9]
- an Output Selector module.

The ability of MISR modules to compact information with a low percentage of aliasing makes them suitable to store the results of the test. In our approach we coupled each module under test with a MISR: whereas the size of the MISR cannot exceed a predefined size, a xor cascade has been used. Each MISR module is reachable from the outside by programming the Output Selector.

This organization allows reducing the re-design operations and supports the reuse of the internal structures: changing ALFSRs and MISRs dimension is a trivial task. Only the individuation of inputs constraints and the design of the Constraints Generator require a bigger effort to designers.

### 3.2. Fault coverage and diagnosis ability evaluation

To obtain high fault coverage and guarantee high ability in terms of fault location, we identified three steps:
1. Statement coverage and toggle activity evaluation
2. Fault coverage measure
3. Equivalent fault classes computation.

In the first step, pseudo-random patterns are applied to the RTL description of the modules composing the logic core and the measure of the percent number of VHDL lines executed is performed. Such measure, usually called statement *coverage*, together with the calculation of the percent number of variables toggled by the patterns, called *toggle activity*, gives to the designer a first degree of confidence about the effectiveness of the generated patterns [10] and can be performed using a simulation tool. Until this step, the evaluated patterns can be generated using the VHDL description of the Pattern Generator or simply calculated with ad-hoc tools generating pseudo-random sequences. Figure 3 represents the first step.

The second step refers to the synthesized component and can be performed using a fault simulator. In order to obtain reliable results, the design to be evaluated in this step should already include the Pattern Generator and the MISRs embedded into the Result Collector, as the final layout optimization will merge their circuitry with that of the device under test.

Whereas the fault coverage reached is less than the required, three actions can be performed:
- apply a larger number of patterns
- modify the ALFSR or MISRs structure
- redefine the Constraints Generator where included.

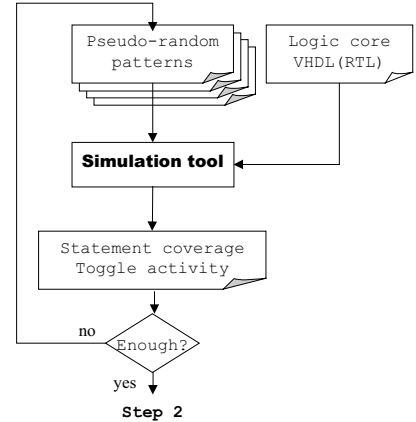

**Fig. 3: Statement coverage and toggle activity evaluation loop.**

While the first action does not require any modification in the designed circuitry, the number of patterns can be increased only until the time for test requirements are not exceeded. In this case, the flow backs to the first step with the evaluation of new patterns. This loop, reported in figure 4, ends when the desired fault coverage is reached or when exceeding the manufacturing constrains.

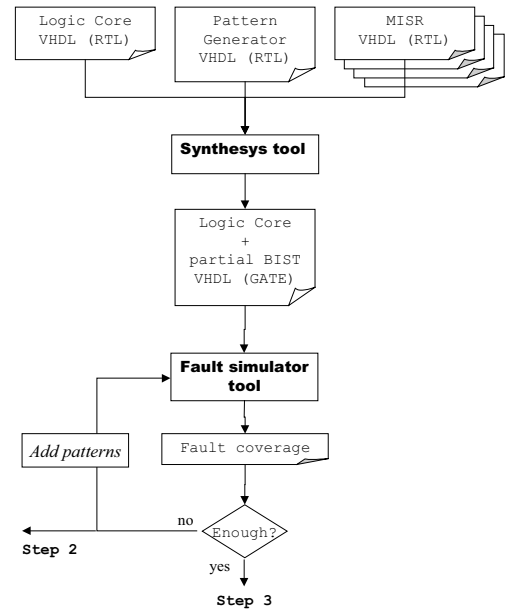

**Fig. 4: Statement coverage and toggle activity evaluation loop.**

The thirst step aims at reaching high diagnosis ability: this characteristic of the performed test can be evaluated by means of the size of the equivalent fault classes [11]. Such measure allows establishing the precision in terms of fault location provided by the analyzed patterns. The size of the equivalent fault classes mainly depends on the ability of the chosen patterns to produce a different *syndrome* for every fault possibly affecting the DUT.

To reach this purpose, a tool able to apply patterns and store the circuit response is needed. The collected information, by means of the obtained syndromes, can be used to build the so-called *diagnostic matrix* [16], allowing to identify the faults belonging to the same equivalent fault class.

To improve the diagnostic properties of the generated patterns, it is possible to operate in two ways:
– adding test patterns
– changing the test structure characteristics.

### 3.3. The P1500 wrapper module

The wrapper, shown in Fig. 5, contains the circuitry necessary to interface the test processor with the outside in a P1500 compliant fashion, supporting the commands for running the BIST operation and accessing to its results. The wrapper is compliant with the suggestions of the P1500 standardization group [12]. The wrapper can be connected to the outside via a standard TAP.

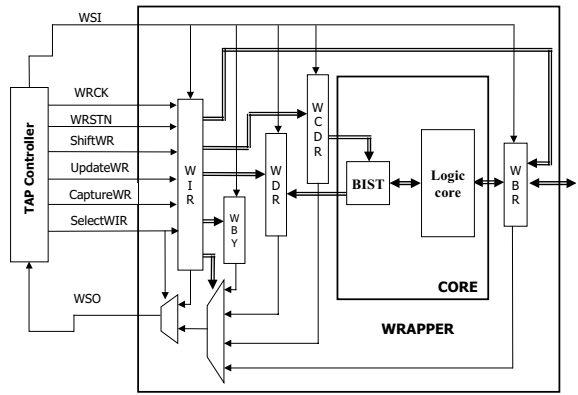

**Fig. 5: Details of the Wrapper Architecture.**

In addition to the mandatory components we propose the introduction of the following Wrapper Data registers:
- *Wrapper Control Data Register* (WCDR): through this register the TAP controller sends the commands to the core (e.g., core reset, core test start, the Status register read, etc.).
- *Wrapper Data Register* (WDR): it is an output register. The TAP Controller can read the test information stored into the status register.

## 4. The case study

We applied the outlined approach to a Reconfigurable Serial Low-Density Parity-Checker decoder core [13-15]. This core was developed by our institution in the frame of a project involving several semiconductor, equipment, and telecom companies; to make it more easily usable by core integrators, a test solution was required: in the following we give details about the adopted solution.

Low-Density Parity-Check (LDPC) codes are powerful and computationally intensive error correction codes, originally proposed by Gallagher [13] and recently rediscovered [14] for a number of applications, including Digital Video Broadcasting (DVB) and magnetic recording. LDPC codes can be represented as a bipartite graph (Fig. 1), where two classes of processing elements iteratively exchange information according to the Message Passing [14-13] algorithm: Bit Nodes (BN) correspond to the codeword symbols, while Check Nodes (CN) are associated to the constraints the code poses on the bit nodes in order for them to form a valid codeword; at each iteration, the reliability of the decoding decisions that can be made on basis of the exchanged information is progressively refined.

Fully and partially parallel solutions for the implementation of the decoder exploit the regularity of the bipartite graph mapping directly CN's and BN's into hardware blocks; graph edges either are mapped to proper interconnect infrastructures or are implemented by means of memory banks. The complexity of check and bit nodes is strongly related to the number of incoming/outgoing edges; additional complexity comes from the requirement of supporting different edge numbers that is typical in the most powerful, irregular codes.

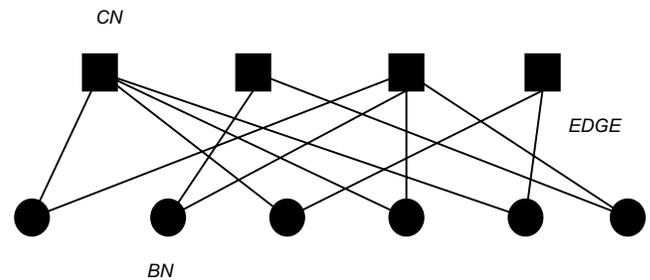

**Fig. 6: Bipartite graph for an LDPC code**

In [15] the implementation of an LDPC decoder is proposed, based on the use of a shared memory that emulates the interconnection between BIT_NODEs and CHECK_NODEs. In [15], a further implementation is proposed, introducing programmability in the architecture proposed in [14]. In this approach, a configurable BIT_NODE and a configurable CHECK_NODE are described. Their ability consists in emulating more than one module, by mapping more "virtual" nodes to the two physically available processing elements; the interconnections are simulated by means of two "interleaving memories" and thanks to its reconfigurable characteristics, this decoder is able to support codes of different sizes and rates, up to a maximum of 512 check nodes and 1,024 bit nodes. A CONTROL UNIT is introduced in order to manage the memory access and the reconfiguration information. The schematic of this enhanced circuitry is reported in Figure 7. Additionally to the use of the showed core in conjunction with external memories to achieve a serial reconfigurable decoder, it can also be adopted as the basic building block for the implementation of a fully parallel architecture.

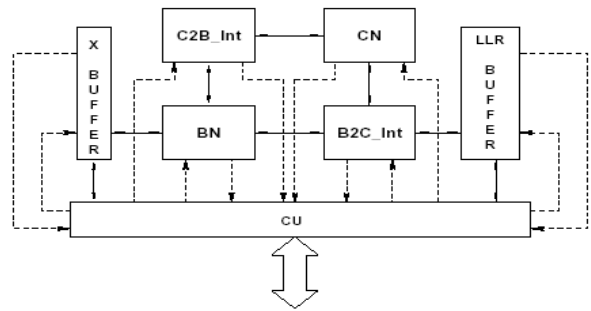

**Fig. 6: Architecture of the Reconfigurable Serial Low Density Parity Checker decoder [15]. The BIT_NODE (BN), the CHECK_NODE (CN) and the CONTROL_UNIT (CU) are connected to the two interleaved memories to perform error detection and correction during transmission of high data volumes**

As far as this design is considered, we partitioned the analyzed logic core in BIT_NODE, CHECK_NODE and CONTROL_UNIT modules. The characteristics of each module in terms of input and output port size is reported in table 1. The test of the two interleaved memories and the buffers is not con-sidered in this paper.

| Component | Input port size [bits] | Output port size [bits] |
|---|---|---|
| BIT_NODE | 54 | 55 |
| CHECK_NODE | 53 | 53 |
| CONTROL_UNIT | 45 | 44 |

**Table 1: Input and output port size in bits.**

The BIST engine has the following structure. The Control Unit contains a counter register (`pattern_counter`) on 12 bits, allowing to apply up to 4,096 patterns for each test execution, and generates a 2 bits signal connected with the Result Collector, in charge of selecting the output to be read.

The Pattern Generator is equipped with a 20-bit ALFRS module and only one Constraints Generator connected both to the BIT_NODE and the CHECK_NODE of the serial LDPC while the CONTROL UNIT does not need it. The Constraints Generator manages a 4 bits sized port that internally selects the data path into the circuitry: it allows applying a limited number of patterns when a small data path is selected, while holding selection values that maximize the used circuitry.

The Result Collector is composed of three 16 bit sized MISR modules, each one connected to the DUT outputs through a xor cascade, and a Output Selector, whose behavior is driven by the Control Unit.

The total area occupied by the DfT additional logic is reported in Tab. 2, and have been worked out by using a commercial tool (*Synopsys Design Analizer*) using an industrial 0.13 μm technological library.

The TAM logic (which includes the Wrapper module) represents a fixed cost necessary to manage the chip-level test. Its area overhead can be quantified as the 16% of the global cost of the additional core-level test logic.

| Component | Area [μm^2] | Overhead [%] |
|---|---|---|
| Serial LDPC | 165,817.88 | - |
| BIST engine | 22,481.63 | 13.5 |
| P1500 Wrapper | 4,566.94 | 2.8 |
| TOTAL | 192,866.51 | 16.4 |

**Table 2: Area overhead evaluation.**

The fault coverage percentage reached by our approach is reported in table 3 and refers to both Stuck At Faults (SAF) and Transition Delay Faults (TDF). Such results have been obtained employing a commercial fault injection tool (Synopsys Tetramax). In order to provide the reader with reference figures, we also included in the table the data related to the cases in which sequential and full scan patterns produced by a commercial ATPG tool (again Synopsys Tetramax) are used. It is important to note that these patterns could not be easily applied to the core, if embedded in a SoC, while the BIST approach is very suitable to deal with this situation. The number of scan cells inserted is 75 for the BIT_NODE, 803 for the CHECK_NODE and 42, divided in two scan chains including 14 and 28 cells, for the CONTROL_UNIT. These values have been calculated using a SUN workstation equipped with a SPARC V8 microprocessor and the CPU times reported in the above table working at 431.03 Mhz in the case of the BIST engine approach (at-speed testing) and at 100 Mhz (supposed ATE frequency) in the Sequential and Full scan approach.

With respect to the Sequential and Full Scan approaches, the use of the BIST approach is desirable for at least the following reasons:

− the fault coverage reached is higher than Sequential patterns coverage and comparable with Full Scan
− the BIST patterns are the same for all modules to be tested, so that they can be tested simultaneously
− the test time is significantly lower for the BIST approach than for the full-scan one
− such patterns are generated and applied one for each clock cycle and results read in the end of the execution, while Sequential and Full scan patterns have to be sent serially by the ATE and results uploaded serially after each operation, thus drastically increasing the ATE storage requirements
− the test patterns are applied by the BIST engine at the nominal frequency of the circuit while the Sequential and Full scan patterns are applied at the ATE frequency that could be lower, guaranteeing more efficiency in the fault coverage.

In table 4, the measure of the performance reduction in terms of frequency lost is reported. This is due to the introduction of the BIST engine and the wrapper. This value is compared with the ones coming from the analysis of the Sequential and Full Scan approach, supposing that:

− for the Sequential approach, patterns are applied using a standard P1500 wrapper
− for the Sequential approach, patterns are applied using a standard P1500 wrapper and introducing into the design multiplexed scan cells.

| Component | | BIST patterns | | Sequential patterns | | Full scan patterns | |
|---|---|---|---|---|---|---|---|
| Fault type | | SAF | TDF | SAF | TDF | SAF | TDF |
| BIT NODE | Faults [#] | 7,532 | 7,532 | 7,532 | 7,532 | 7,836 | 7,836 |
| | FC [%] | 97.8 | 95.6 | 93.8 | 84.3 | 98.5 | 91.2 |
| | clock cycles | 4,096 | 4,096 | 11,340 | 16,580 | 21,248 | 39,168 |
| | CPU time | - | - | 489 sec | 2,628 sec | 197 sec | 277 sec |
| CHECK NODE | Faults [#] | 86,104 | 86,104 | 86,104 | 86,104 | 89,412 | 89,412 |
| | FC [%] | 91.6 | 90.7 | 82.9 | 76.4 | 93.1 | 87.1 |
| | Clock cycles | 4,096 | 4,096 | 8374 | 7844 | 380,064 | 866,272 |
| | CPU Time | - | - | ~ 54 h | ~ 43 h | 428 sec | 692 sec |
| CONTROL UNIT | Faults [#] | 3,038 | 3,038 | 3,038 | 3,038 | 3,216 | 3,216 |
| | FC [%] | 97.5 | 95.3 | 89.8 | 84.0 | 98.6 | 91.3 |
| | Clock cycles | 4,096 | 4,096 | 3060 | 4,860 | 16,965 | 27,405 |
| | CPU time | - | - | 2422 sec | 5909 sec | 91 sec | 123 sec |

**Table 3: Fault coverage.**

|  | Original design | BIST engine | Sequential approach | Full scan approach |
|---|---|---|---|---|
| frequency [MHz] | 438.6 | 431.03 | 434.14 | 426.62 |

**Table 4: Performance reduction for the investigated approaches.**

Finally, table 5 shows the size of the equivalent fault classes for the three components obtained for the BIST engine, Sequential patterns and Full Scan approach applying the number of patterns reported in table 2. That result have been obtained exploiting an in-home developed tool in C language: this tool is able to build and analyze the diagnostic matrix by collecting each fault syndromes obtained by using *Synopsys Tetramax* as a fault simulator.

| Component | BIST patterns | | Sequential patterns | | Full scan Patterns | |
|---|---|---|---|---|---|---|
|  | Max size | Med size | Max size | Med size | Max size | Med size |
| BIT_NODE | 3 | 1.2 | 7 | 4.4 | 3 | 1.6 |
| CHECK_NODE | 4 | 1.9 | 12 | 6.9 | 7 | 2.7 |
| CONTROL_UNIT | 2 | 1.3 | 8 | 5.1 | 2 | 1.3 |

**Table 5: Equivalent fault classes maximum and medium size obtained by the investigated approach.**

## 5. Conclusions

In this paper we presented a case study in which a logic core for telecom applications was equipped with test features suitable to guarantee a high fault coverage (especially with respect to delay faults) and a high modularity and easy integrability into a SoC, even from the point of view of testing.

The adopted approach is based on a BIST engine that is in charge of applying test patterns and observing the module behavior. Such BIST engine has been designed to allow high flexibility in order to easily adopt the same approach to a large set of different core models and easily allow the upload of test results. Thanks to the adopted solution, the test can be applied at-speed, thus reaching a very high fault coverage with respect to delay faults. The proposed architecture exploits the current test standards interface (IEEE 1149.1 and P1500 standards) in order to simplify the design effort. Moreover, the P1500 wrapper eases the integration of the core into the overall SoC test strategy. The test architecture is transparent to the core user, thus guaranteeing the protection of the intellectual property.